\documentclass[12pt,fleqn]{iopart}
\usepackage{iopams}
\usepackage[dvips]{graphicx}
\usepackage{subfigure}
\usepackage{bm,amssymb}

\begin{document}
\title{Magneto-orbital coupling in iron pnictides.}
\author{Sayandip Ghosh, Nimisha Raghuvanshi, and Avinash Singh}
\address{Department of Physics, Indian Institute of Technology Kanpur
208016, India} 
\ead{sayandip@iitk.ac.in}
\date{\today} 

\begin{abstract}
A magneto-orbital coupling mechanism is proposed to account for the weak energy gap at the Fermi energy in the $(\pi,0)$ ordered SDW state of a realistic three band model for iron pnictides involving $d_{xz}$, $d_{yz}$, and $d_{xy}$ Fe orbitals. The orbital mixing terms between the $d_{xy}$ and $d_{xz}/d_{yz}$ orbitals, which are important in reproducing the orbital composition of the elliptical electron pockets at $(\pm \pi,0)$ and $(0,\pm \pi)$, are shown to play a key role in the energy gap formation in the SDW state.
\end{abstract}
\maketitle

\section{Introduction}
The iron pnictides exhibit a rich phase diagram
\cite{Zhao2008nature,Nandi2010} including magnetic, structural
(tetragonal-to-orthorhombic), and superconducting phase
transitions \cite{Fernandes2014}. The magnetic state exhibits
$(\pi,0)$ ordering of Fe moments in the $a$-$b$ plane, with a concomitant structural distortion $a$>$b$ possibly correlated with the ferro orbital order $d_{xz}$$>$$d_{yz}$ as seen in ARPES studies \cite{Yi2011}. Inelastic neutron scattering experiments \cite{Ewings2011,Harriger2011} have shown well defined spin wave
excitations with energy scale $\sim$ 200 meV, persisting even above the N\'{e}el temperature \cite{Lu2014}. Evidently, short range antiferromagnetic (AF) and ferromagnetic (F) order remain in the $a$ and $b$ directions respectively even above the disordering temperature for long-range magnetic order, which may account for the narrow nematic phase \cite{Kasahara2013,Zhou2013} above the N\'{e}el temperature where the ferro orbital order \cite{Yi2011}
and structural distortion survive, as well as the temperature dependence of the measured anisotropies in $a$ and $b$ directions of magnetic excitations and resistivity \cite{Yi2011,Lu2014,Chu2010}.

Evidently, the complex multi-orbital character of the underlying
microscopic description of iron pnictides plays an important role in understanding their macroscopic behavior as seen from the interplay of magnetic, orbital, structural, and transport properties. Another important macroscopic behavior of iron pnictides is associated with the energy gap in the $(\pi,0)$ ordered magnetic state, as seen in optical conductivity, transport and scanning tunneling microscopy studies; investigating the role of the multi-orbital character on the magnetic-state electronic structure is therefore of interest.

For the non-magnetic state, intensive investigations of the electronic structure of these materials by first-principle calculations \cite{Mazin2008,Haule2008,Singh2008,Nekrasov2008,Zhang2009} and angle resolved photoemission spectroscopy (ARPES) experiments \cite{Yi2009,Kondo2010,Yi2011,Brouet2012,Kordyuk2013} have revealed that the Fermi surface consists of two nearly circular hole pockets around the center and  elliptical electron pockets around the corner of the BZ which are primarily contributed by $d_{xz}$, $d_{yz}$, and $d_{xy}$ Fe $3d$ orbitals.

A variety of experimental techniques have confirmed the opening of a small energy gap in the excitation spectrum associated with the onset of the $(\pi,0)$ spin-density–wave (SDW) order in iron pnictides \cite{Johnston2010}. The Fermi surface (FS) undergoes a complex multi-orbital reconstruction through the non-magnetic to antiferromagnetic transition. ARPES experiments have revealed that the electronic states are strongly modified across the SDW transition \cite{Jong2010,Zhou2010,Yang2009}. Bands get folded and hybridized, resulting in band splitting and opening of SDW gaps. Consequently, the Fermi surface breaks up into small droplet-like structures, supporting only a fraction of the original FS areas. These results are consistent with Quantum Oscillation measurements \cite{Sebastian2008,Analytis2009} showing the reconstruction and drastic reduction of Fermi surface area, which can be as small as $\sim 1 \%$ of the original BZ area for $\rm SrFe_2As_2$. 

In addition, investigations of optical and transport properties by optical spectroscopy \cite{Hu2008,Wu2009} and broadband spectroscopic ellipsometry \cite{Charnukha2013} measurements on single-crystalline iron pnictides have found an energy gap for electronic excitations in the SDW ordered state. The carrier density is reduced substantially in the ordered state
with consequent decrease in optical conductivity. Also, as inferred from the results of infrared studies, short-range AF correlations have been suggested as being responsible for the pseudogap in iron superconductors \cite{Moon2014,Moon2012}.
Moreover, scanning tunneling microscopy (STM) studies have also seen an energy gap in the density of states near the Fermi energy \cite{Zhang2010,Dutta2013}.

Close proximity of the superconducting phase and magnetically ordered parent phase in iron pnictides have naturally led to comparisons with cuprates \cite{Moon2014}.
However, the magnitude of the experimentally measured energy gap at the Fermi energy (up to $\sim 150$ meV for $\rm Ca Fe_2 As_2$) is significantly smaller than the usual SDW gap $2 \Delta = mU$ associated with AF ordering with staggered magnetization $m$ and local Coulomb interaction $U$, as in the Hubbard model representation for cuprates. Moreover, these SDW gaps open at energies typically away from the Fermi energy in realistic multi-band tight binding models which yield the correct FS structure in agreement with DFT calculations and ARPES studies in the non-magnetic state.

In this paper, a magneto-orbital coupling mechanism is proposed  for a realistic three-band model at half filling for iron pnictides involving $d_{xz}$, $d_{yz}$, and $d_{xy}$ orbitals of Fe, which is shown to account for the weak non-conventional energy gap at the Fermi energy due to a composite effect of the orbital mixing (hybridization) between the $d_{xy}$ and $d_{xz}/d_{yz}$ orbitals and the $(\pi,0)$ SDW magnetic order. For realistic three-band-model parameters, the energy gap is found to be in quantitative agreement with experiments. The gapped SDW state in the realistic three band model at half filling therefore provides, in analogy with cuprates, a suitable reference state for theoretically investigating effects of electron/hole doping in iron pnictides.

\section{Three-orbital model and ($\pi,0$) SDW state}
We consider a minimal three-orbital model \cite{Ghosh2015} involving $d_{xz}$, $d_{yz}$ and $d_{xy}$ Fe $3d$ orbitals. The tight binding Hamiltonian in the plane-wave basis is defined as:
\begin{equation}
 H_0 = \sum_{\bf k} \sum_{\sigma} \sum_{\mu,\nu} T^{\mu,\nu}({\bf k}) a_{{\bf
k},\mu,\sigma}^{\dagger} a_{{\bf k},\nu,\sigma},
\label{tight}
\end{equation}
where
\begin{eqnarray}
T^{11} &=& - 2t_1\cos  k_x - 2t_2\cos  k_y - 4t_3 \cos  k_x 
\cos  k_y \label{eq:t11}\nonumber \\
T^{22} &=& - 2t_2\cos  k_x -2t_1\cos  k_y - 4t_3 \cos  k_x 
\cos  k_y \label{eq:t22}\nonumber \\
T^{33} &=& - 2t_5(\cos  k_x+\cos  k_y)  - 4t_6\cos  k_x\cos  k_y 
+ \varepsilon_{\rm diff}  \label{eq:t33} \nonumber \\
T^{12} &=& T^{21} =- 4t_4\sin  k_x \sin  k_y \label{eq:t12}\nonumber \\
T^{13} &=& \bar{T}^{31} = - 2it_7\sin  k_x - 4it_8\sin  k_x \cos  k_y
\label{eq:t13} \nonumber \\
T^{23} &=& \bar{T}^{32}= - 2it_7\sin  k_y - 4it_8\sin  k_y \cos  k_x\;
\label{eq:t23}
\end{eqnarray}
are the tight-binding matrix elements in the unfolded BZ ($-\pi \leq k_x,k_y \leq \pi$). Here, $t_1$ and $t_2$ are the intra-orbital hoppings for $xz$ ($yz$) along $x$ ($y$) and $y$ ($x$) directions, respectively, $t_3$ and $t_4$ are the intra and inter-orbital hoppings along diagonal direction for $xz$ and $yz$, $t_5$ and $t_6$ are intra-orbital NN and NNN hoppings for $xy$, while $t_7$ and $t_8$ the NN and NNN hybridization between $xy$ and $xz/yz$. Finally, $\varepsilon_{\rm diff}$ is the energy difference between the $xy$ and degenerate $xz/yz$ orbitals.

\begin{figure}
\begin{center}
 \subfigure[]{\includegraphics[height=60mm,angle=-90]{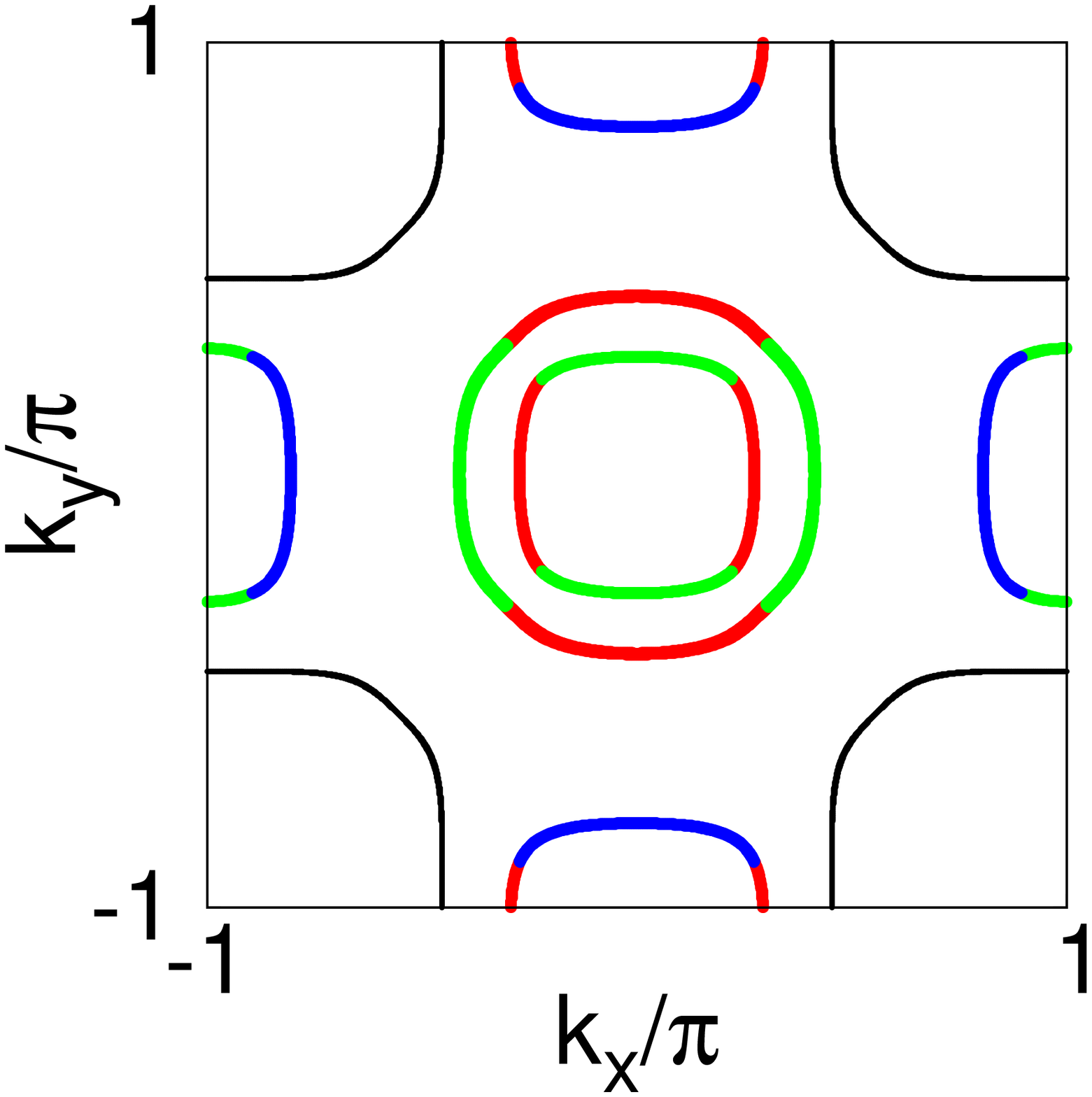}
 \label{fs}}\\
\subfigure[]{\includegraphics[height=50mm,angle=0]{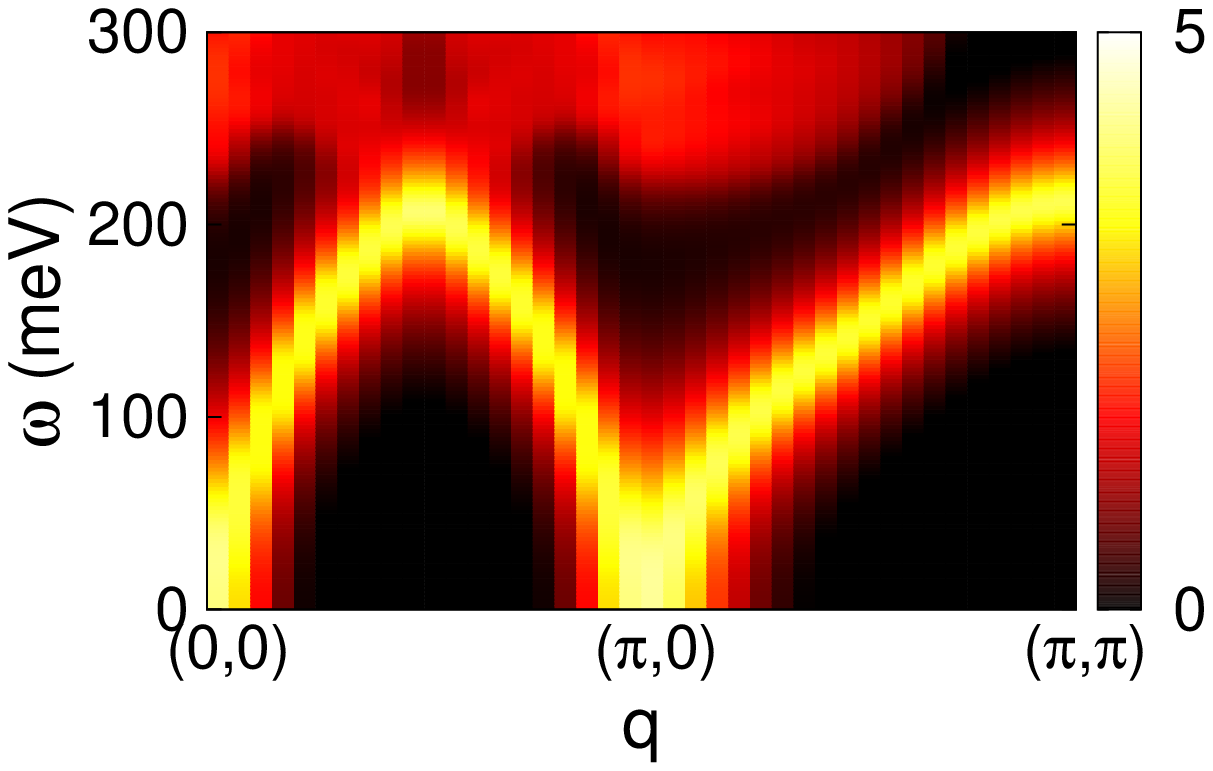}
 \label{sw}}
 \caption{ \label{model}\subref{fs} Fermi surface in the unfolded BZ for the three-orbital model with hopping parameters as given in Table \ref{hoppings}. The main orbital contributions are shown as: $d_{xz}$ (red), $d_{yz}$ (green), and $d_{xy}$ (blue). \subref{sw} The spin wave spectral function in the ($\pi,0$) SDW state for the three-orbital model at half filling.}
 \label{fs}
\end{center}
\end{figure}

 \begin{table}
\caption{Values of the hopping parameters in the three-orbital model (in eV)} 
\begin{center}
\addtolength{\tabcolsep}{5pt}
 \begin{tabular}{c c c c c c c c c}\hline 
$t_1$ & $t_2$ & $t_3$ & $t_4$ & $t_5$ & $t_6$ & $t_7$ & $t_8$ \\ \hline
 $0.1$ & $0.32$ & $-0.29$ & $-0.06$ & $-0.3$ & $-0.16$ & $-0.15$ & $-0.02$ \\ \hline 
 \end{tabular}
\end{center}
 \label{hoppings}
\end{table}

The Fermi surface for the three-orbital Hamiltonian (\ref{tight}) for the values of hopping parameters given in Table 1 is shown in Figure 1(a). Here, the Fermi energy is kept at $-0.06$ eV corresponding to half filling. There are two circular hole pockets around the center and elliptical electron pockets around ($\pm \pi,0$) and ($0,\pm \pi$) in the unfolded BZ. The two hole pockets involve primarily the $xz$ and $yz$ orbitals, while the electron pockets centered at ($\pm \pi,0$) [($0,\pm \pi$)] arise mainly from the hybridization of the $xy$ and $yz$ [$xz$] orbitals. All of these features are in good agreement with results from DFT calculations and ARPES experiments. 

We now consider the ($\pi,0$) ordered magnetic (SDW) state of this model. The various electron-electron interaction terms  included are:

\begin{eqnarray}
H_I &=& U \sum_{{\bf i},\mu} n_{{\bf i},\mu,\uparrow} n_{{\bf i},\mu,\downarrow} + (U' -
\frac{J}{2}) \sum_{{\bf i},\mu,\nu}^{\mu<\nu} n_{{\bf i},\mu} n_{{\bf i},\nu} - 2 J \sum_{{\bf i},\mu,\nu}^{\mu<\nu} {\bf{S_{{\bf i},\mu}}} \cdot {\bf{S_{{\bf i},\nu}}} \nonumber \\
&+& J' \sum_{{\bf i},\mu,\nu}^{\mu<\nu}
(a_{{\bf i},\mu,\uparrow}^{\dagger}a_{{\bf i},\mu,\downarrow}^{\dagger}a_{{\bf i},\nu,\downarrow}
a_{{\bf i},\nu,\uparrow} + \rm{H.c.}),
\label{interaction}
\end{eqnarray}
where ${\bf S}_{i,\mu}$ ($n_{{\bf i},\mu}$) refer to the local spin (charge) density operators for orbital $\mu$. The first and second terms are the intra-orbital and inter-orbital Coulomb interactions respectively, the third term is the Hund's coupling and the fourth term the ``pair-hopping'' term.

Extending the two-sublattice basis approach for the SDW state in a single-band model \cite{Raghuvanshi2010} to a composite three-orbital, two-sublattice basis, the Hartree-Fock (HF) level Hamiltonian matrix in this composite basis (A$xz$ A$yz$ A$xy$ B$xz$ B$yz$ B$xy$) is obtained as:

\begin{footnotesize}
\begin{eqnarray}
\fl
H_{\rm HF}^{\sigma} ({\bf k}) =  
\left [ \begin{array}{cccccc} -\sigma
\Delta_{xz} + \varepsilon_{\bf k}^{2y}   & 0 & 0 & \varepsilon_{\bf k}^{1x}
+ \varepsilon_{\bf k}^{3} & \varepsilon_{\bf k}^{4} & \varepsilon_{\bf k}^{7x}
+ \varepsilon_{\bf k}^{8,1}
\\ 
0 & -\sigma \Delta_{yz} + \varepsilon_{\bf k}^{1y} & \varepsilon_{\bf k}^{7y}  &
\varepsilon_{\bf k}^{4} & \varepsilon_{\bf k}^{2x} + \varepsilon_{\bf k}^{3}  &
\varepsilon_{\bf k}^{8,2} \\ 
0 & -\varepsilon_{\bf k}^{7y} & - \sigma \Delta_{xy} + \varepsilon_{\bf
k}^{5y} + \varepsilon_{\rm diff} & - \varepsilon_{\bf k}^{7x}- \varepsilon_{\bf
k}^{8,1} & -\varepsilon_{\bf k}^{8,2} & \varepsilon_{\bf k}^{5x} +
\varepsilon_{\bf k}^{6} \\ 
\varepsilon_{\bf k}^{1x} + \varepsilon_{\bf k}^{3} & \varepsilon_{\bf k}^{4} &
\varepsilon_{\bf k}^{7x} + \varepsilon_{\bf k}^{8,1} & \sigma \Delta_{xz} +
\varepsilon_{\bf k}^{2y} & 0 & 0 \\
\varepsilon_{\bf k}^{4} & \varepsilon_{\bf k}^{2x} + \varepsilon_{\bf k}^{3} &
\varepsilon_{\bf k}^{8,2} & 0 & \sigma \Delta_{yz} + \varepsilon_{\bf k}^{1y} &
\varepsilon_{\bf k}^{7y} \\
- \varepsilon_{\bf k}^{7x} - \varepsilon_{\bf k}^{8,1} & -\varepsilon_{\bf
k}^{8,2} & \varepsilon_{\bf k}^{5x} + \varepsilon_{\bf k}^{6} & 0 &
- \varepsilon_{\bf k}^{7y} & \sigma
\Delta_{xy} + \varepsilon_{\bf k}^{5y} + \varepsilon_{\rm diff}  \\
\end{array}
\right ] \nonumber \\ 
\label{Hamiltonian}
\end{eqnarray}
\end{footnotesize}
for spin $\sigma$, where
\begin{eqnarray}
\varepsilon_{\bf k}^{1x} &=& -2 t_1 \cos k_x  \;\;\;\;\;\;
\varepsilon_{\bf k}^{1y} = -2 t_1 \cos k_y  \nonumber \\
\varepsilon_{\bf k}^{2x} &=& -2 t_2 \cos k_x  \;\;\;\;\;\; 
\varepsilon_{\bf k}^{2y} = -2 t_2 \cos k_y  \nonumber \\
\varepsilon_{\bf k}^{5x} &=& -2 t_5 \cos k_x  \;\;\;\;\;\; 
\varepsilon_{\bf k}^{5y} = -2 t_5 \cos k_y  \nonumber \\
\varepsilon_{\bf k}^{3} &=& -4 t_3 \cos k_x \cos k_y  \;\;\;\;\;\;
\varepsilon_{\bf k}^{4} = -4 t_4 \sin k_x \sin k_y \nonumber \\ 
\varepsilon_{\bf k}^{6} &=& -4 t_6 \cos k_x \cos k_y \nonumber \\
\varepsilon_{\bf k}^{7x} &=& -2i t_7 \sin k_x  \;\;\;\;\;\;
\varepsilon_{\bf k}^{7y} = -2i t_7 \sin k_y  \nonumber \\
\varepsilon_{\bf k}^{8,1} &=& -4i t_8 \sin k_x \cos k_y  \;\;\;\;\;\;
\varepsilon_{\bf k}^{8,2} = -4i t_8 \cos k_x \sin k_y \nonumber \\ 
\end{eqnarray}
are the band energies corresponding to the hopping terms along different directions, and the self-consistent exchange fields are defined as $2\Delta_\mu = U m_{\mu} + J\sum_{\nu \neq \mu}m_{\nu}$ in terms of sublattice magnetization $m_{\mu}$ for orbital $\mu$.  

The calculated spin wave spectral function \cite{Ghosh2015} in the SDW state for the three-orbital model is shown in Fig. 1(b). Evidently, spin wave excitations are highly dispersive, and do not decay into the particle-hole continuum. The energy scale of spin excitations is $\sim$ 200 meV with a well-defined maximum at the ferromagnetic zone boundary [${\bm Q}=(\pi,\pi)$]. These features of spin wave excitations are in excellent agreement with results from inelastic neutron scattering measurements, confirming that our three-orbital model is realistic.

\section{Electronic structure in the SDW state}

Figure 2 shows the the evolution of the orbital resolved density of states (DOS) near the Fermi energy in the SDW state with increasing interaction strength $U$. The corresponding band structures along symmetry directions of the BZ are shown in Fig. 3. Figures 2(a) and 3(a) correspond to the PM state for which the model yields correct FS structure as shown in Fig. 1(a). 

As seen in Fig. 2, a large gap opens for the $yz$ band in the SDW state. This band gap is a conventional SDW gap, which opens at wave vector $k_x = \pi/2$ in the AF direction, as seen from the band dispersion plot shown in Fig. 3. Similar SDW gaps open up for the $xz$ and $xy$ bands at $k_x = \pi/2$. However, these band gaps appear at energies far away from the Fermi energy. Besides these three conventional SDW band gaps, there is a weak band gap involving the $xy$ and $xz$ orbitals which opens up near the Fermi energy, as seen in Fig. 3, which is found to be robust with respect to small variations in the hopping parameters.

It is important to note that while partial density of states for $xz$ and $yz$ orbitals are symmetric in the PM state [Fig. 2(a)], they become anisotropic in the SDW state [2(b)]. The DOS for the $yz$ orbital has a peak further away from Fermi energy ($E_F$) than that for $xz$ and has a relatively smaller contribution at $E_F$. This is in excellent agreement with results of DMFT studies \cite{Yin2011} and provides a good understanding of the ferro orbital ordering between the two orbitals \cite{Kim2013,Ghosh2014}. Moreover, the band dispersion, gap formation, as well as the largest gap for $yz$ [Fig. 2(c) and 2(d)] agree well with the self-consistent calculations for multi-orbital models \cite{Plonka2013,Yi2014}. 

Physically, this weak energy gap near the Fermi energy involving the $xy$ and $xz$ orbitals is the most important. One of the key contributing factors for this energy gap formation is the orbital mixing between the $xz$ and $xy$ orbitals due to the hopping terms $t_7$ and $t_8$. Only $t_7$ will be retained in the discussion below as $t_8$ is negligible in comparison (see Table 1). Fig. 4 shows the SDW state band structure with and without $t_7$, confirming the role of the orbital mixing term in the gap formation. It is important to note that the opening of this gap for the $xz$ and $xy$ orbitals due to the orbital mixing term $t_7$ critically requires the presence of the SDW magnetic ordering, as is evident from the absence of any gap in the non-magnetic state ($\Delta =0$), even with the mixing term $t_7$ included [Fig. 4 (a)]. For simplicity, we have taken equal exchange fields ($\Delta=0.3$ eV, $U\approx 1$ eV) for all three orbitals.

To summarize, {\it both} the SDW order as well as the orbital mixing term $t_7$ between the $xz$ and $xy$ orbitals are critically important in the formation of the weak energy gap near the Fermi energy. Only the $t_7$ term (non-magnetic state) or only the SDW order ($t_7= 0$) are independently not sufficient to open the gap, both must be present together. 

\section{Magneto-orbital coupling}

A magneto-orbital coupling mechanism is proposed below to account for this non-conventional weak energy gap near the Fermi energy involving the $xz$ and $xy$ orbitals. In the ($\pi,0$) ordered SDW state with AF ordering in the $x$ direction, the orbital mixing term  $2i (t_7 + 2t_8) \sin k_x$ connects $xz$ and $xy$ orbitals on opposite sublattices, highlighting the coupling between the 
orbital and magnetic sectors.

\begin{figure}
\begin{center}
 \subfigure{\includegraphics[height=50mm]{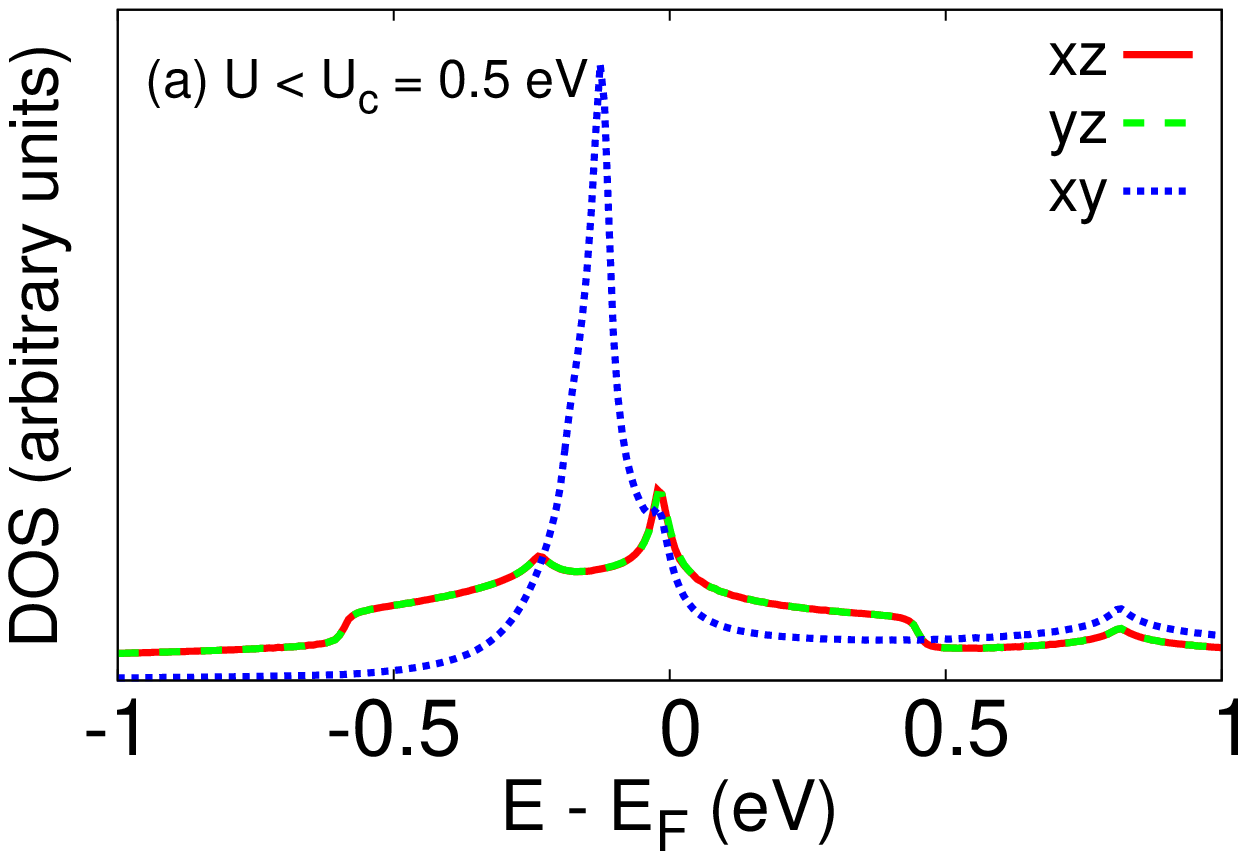}
 \label{dos1}}
\hspace{+0.0cm}
 \subfigure{\includegraphics[height=50mm]{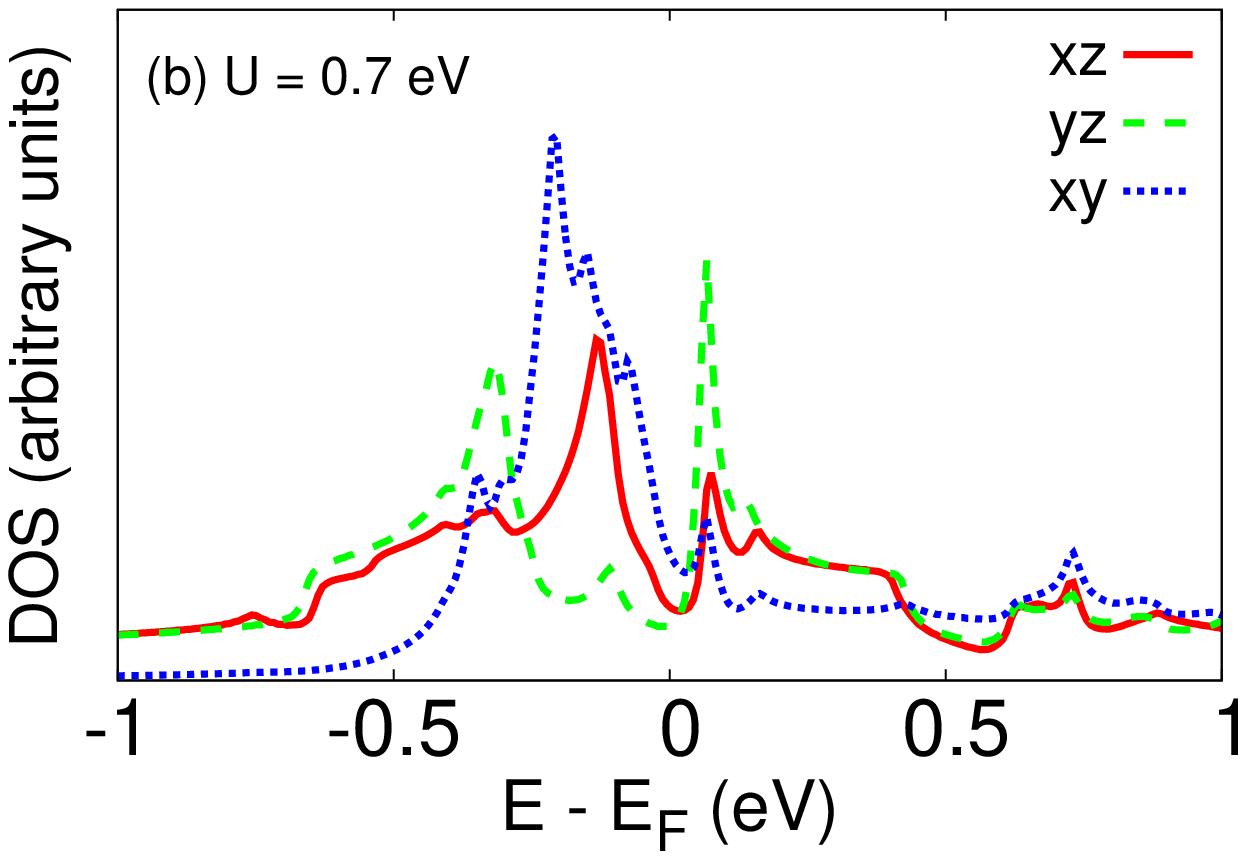}
  \label{dos2}}\\
\vspace{+0.0cm}
 \subfigure{\includegraphics[height=50mm]{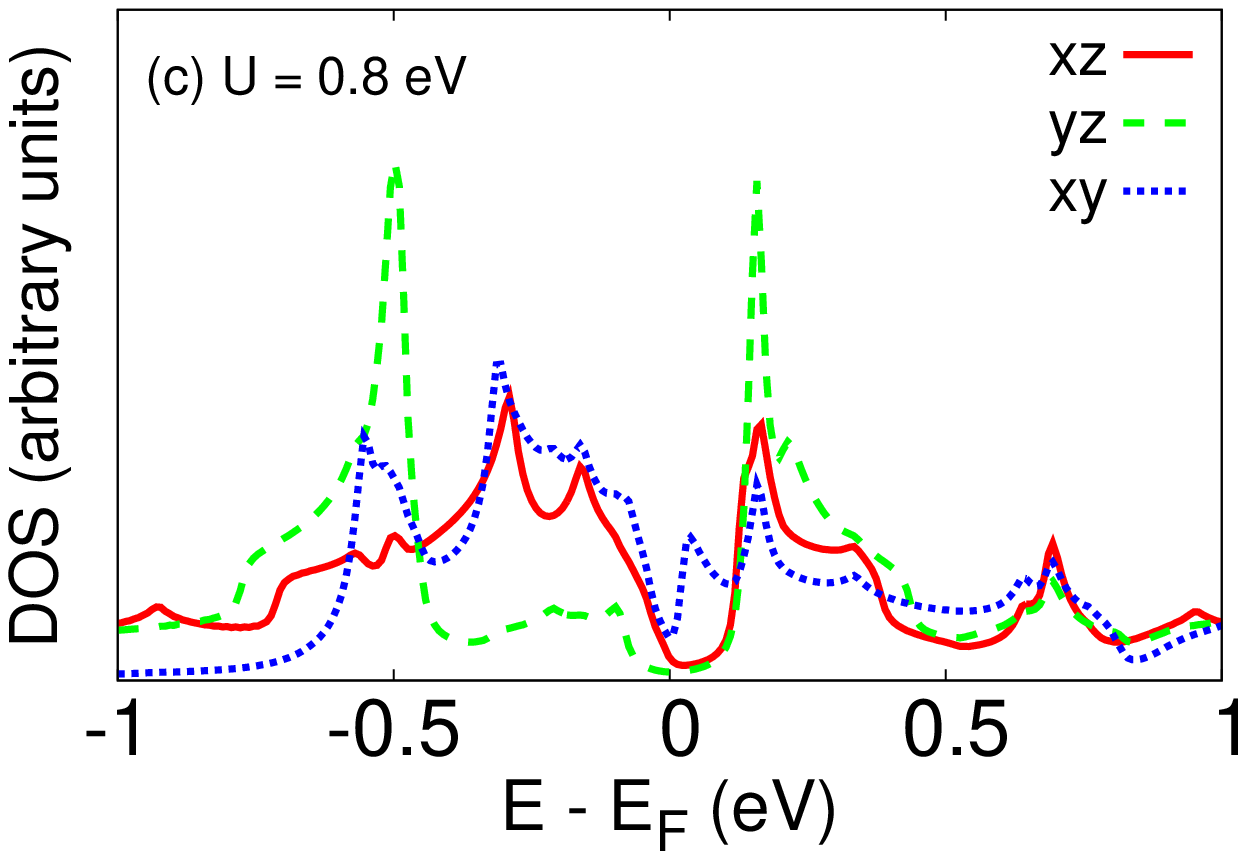}
 \label{dos3}}
\hspace{+0.0cm}
 \subfigure{\includegraphics[height=50mm]{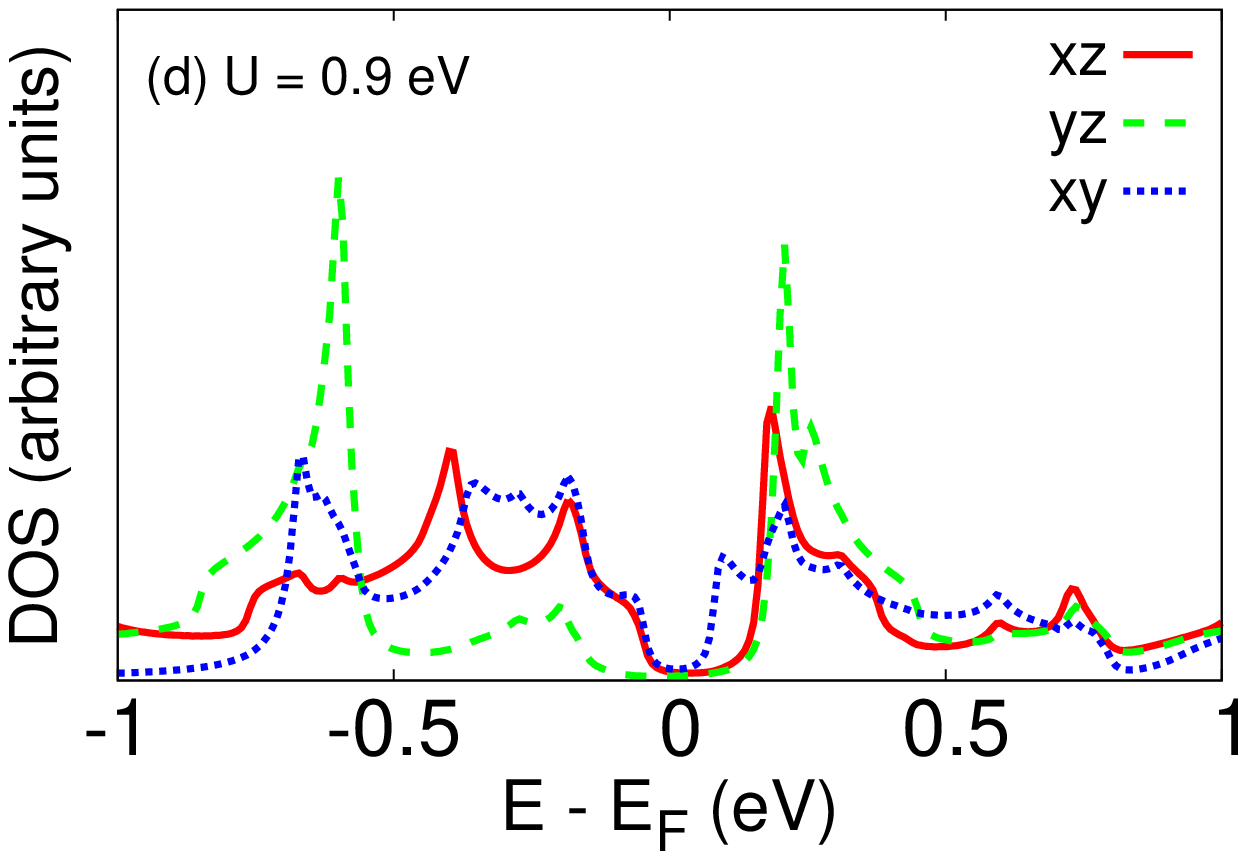}
  \label{dos4}}
 \caption{Evolution of the orbitally resolved DOS in the $(\pi,0)$ SDW state of the three-band model, showing the opening of the energy gap with increasing interaction strength $U$ at the Fermi energy corresponding to half filling.}
 \label{dos}
\end{center}
\end{figure}

\begin{figure}
\begin{center}
 \subfigure{\includegraphics[height=50mm]{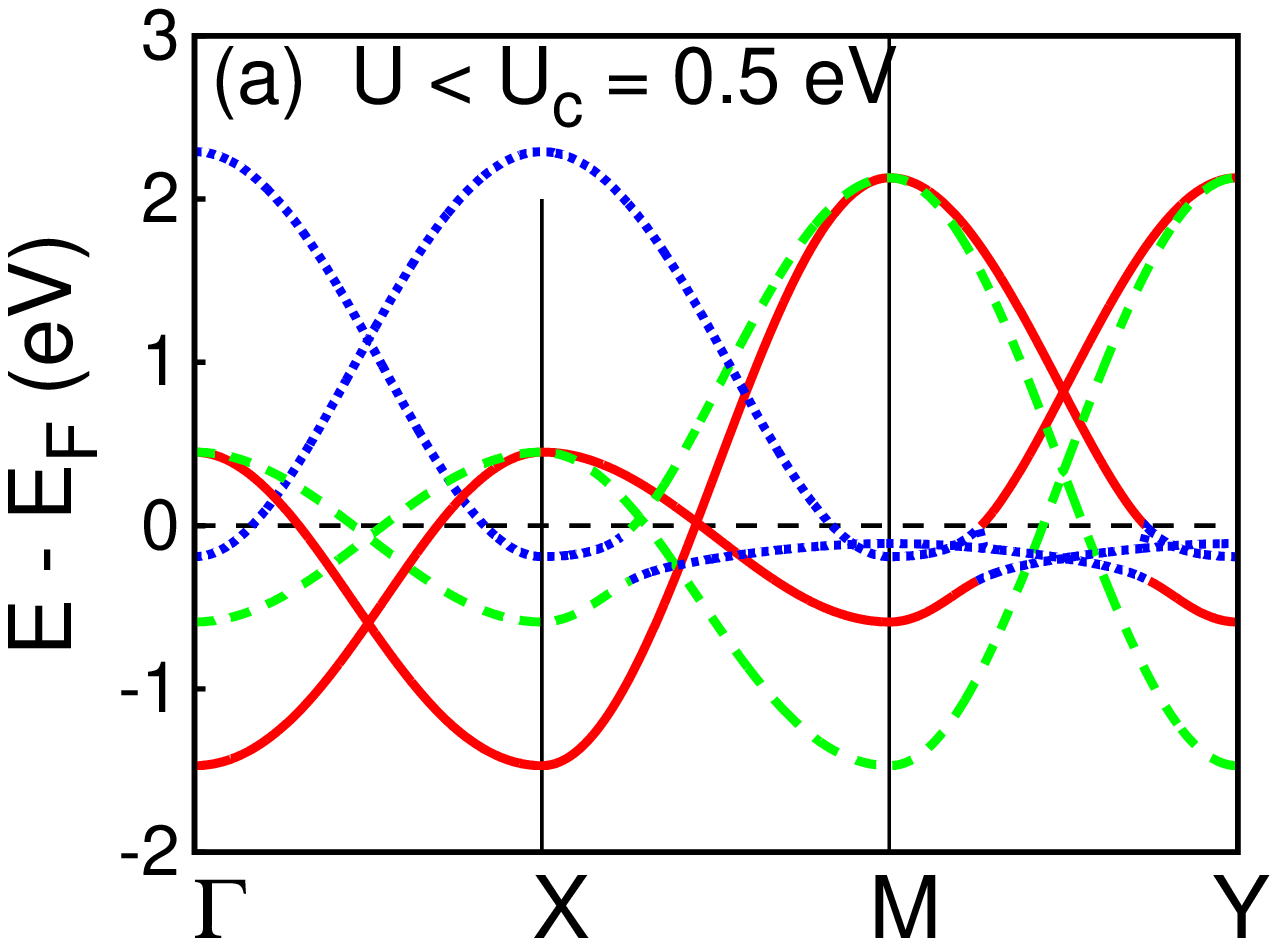}
 \label{dos1}}
\hspace{+0.0cm}
 \subfigure{\includegraphics[height=50mm]{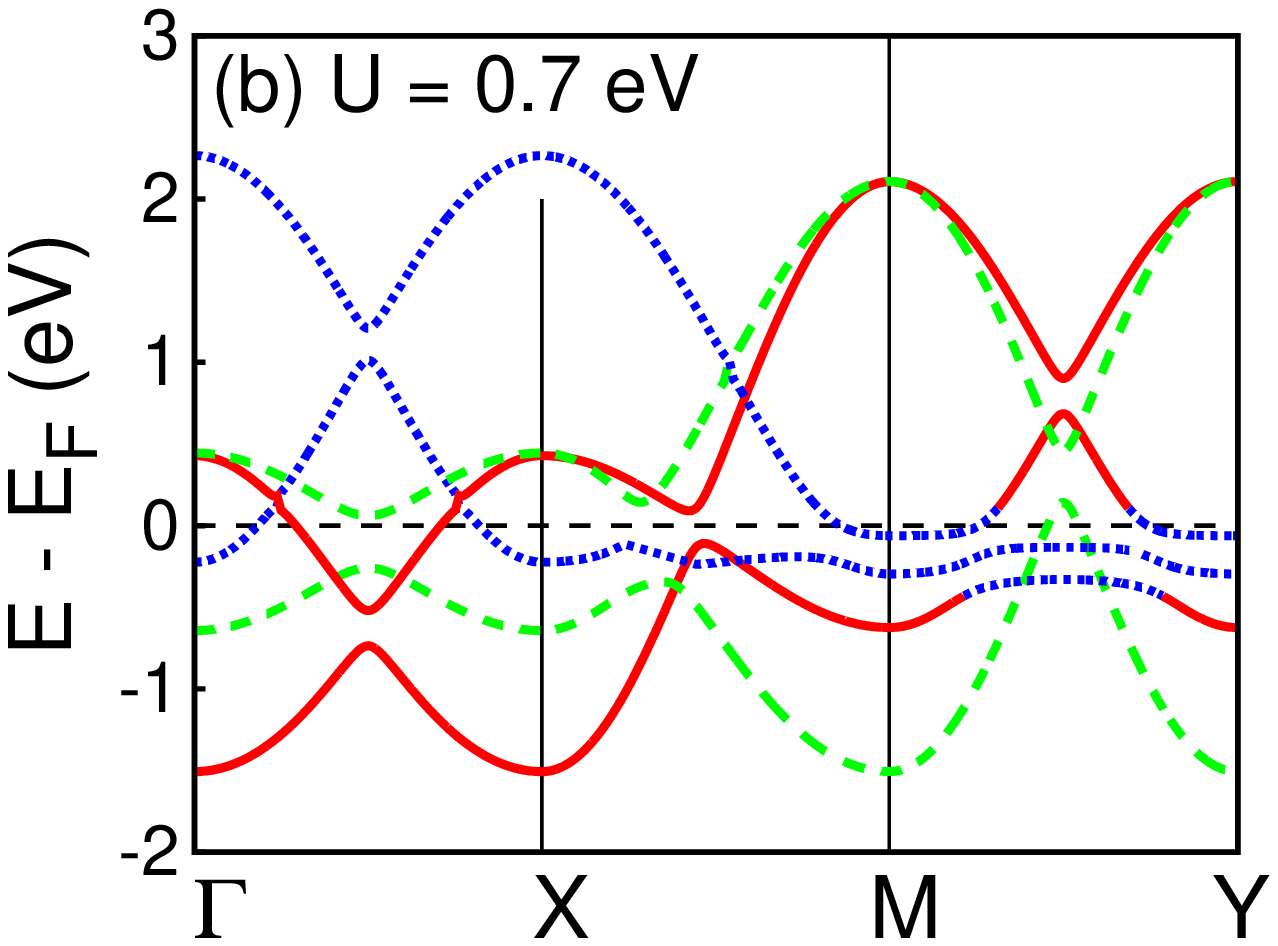}
  \label{dos2}}\\
\vspace{+0.0cm}
 \subfigure{\includegraphics[height=50mm]{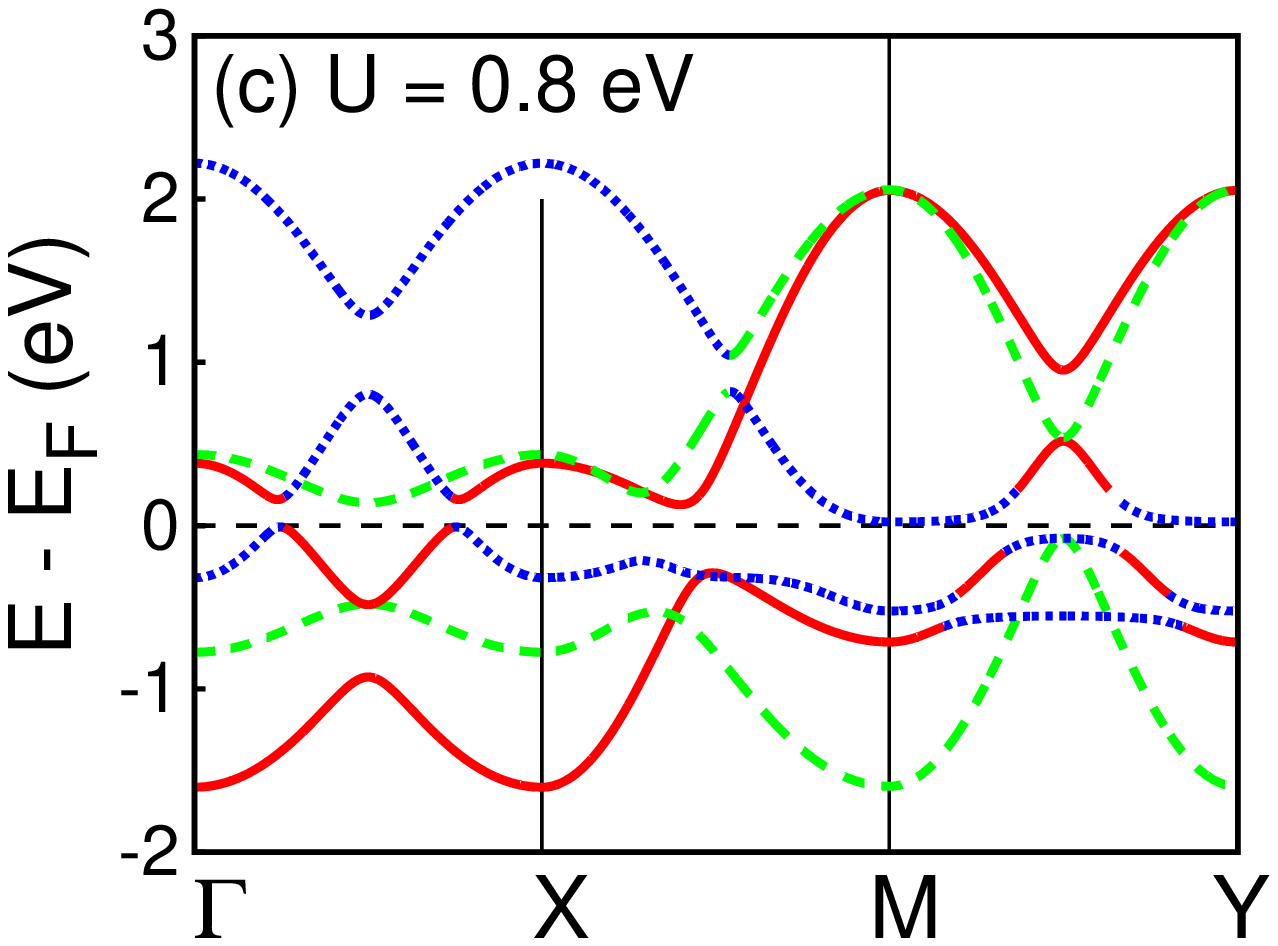}
 \label{dos3}}
\hspace{+0.0cm}
 \subfigure{\includegraphics[height=50mm]{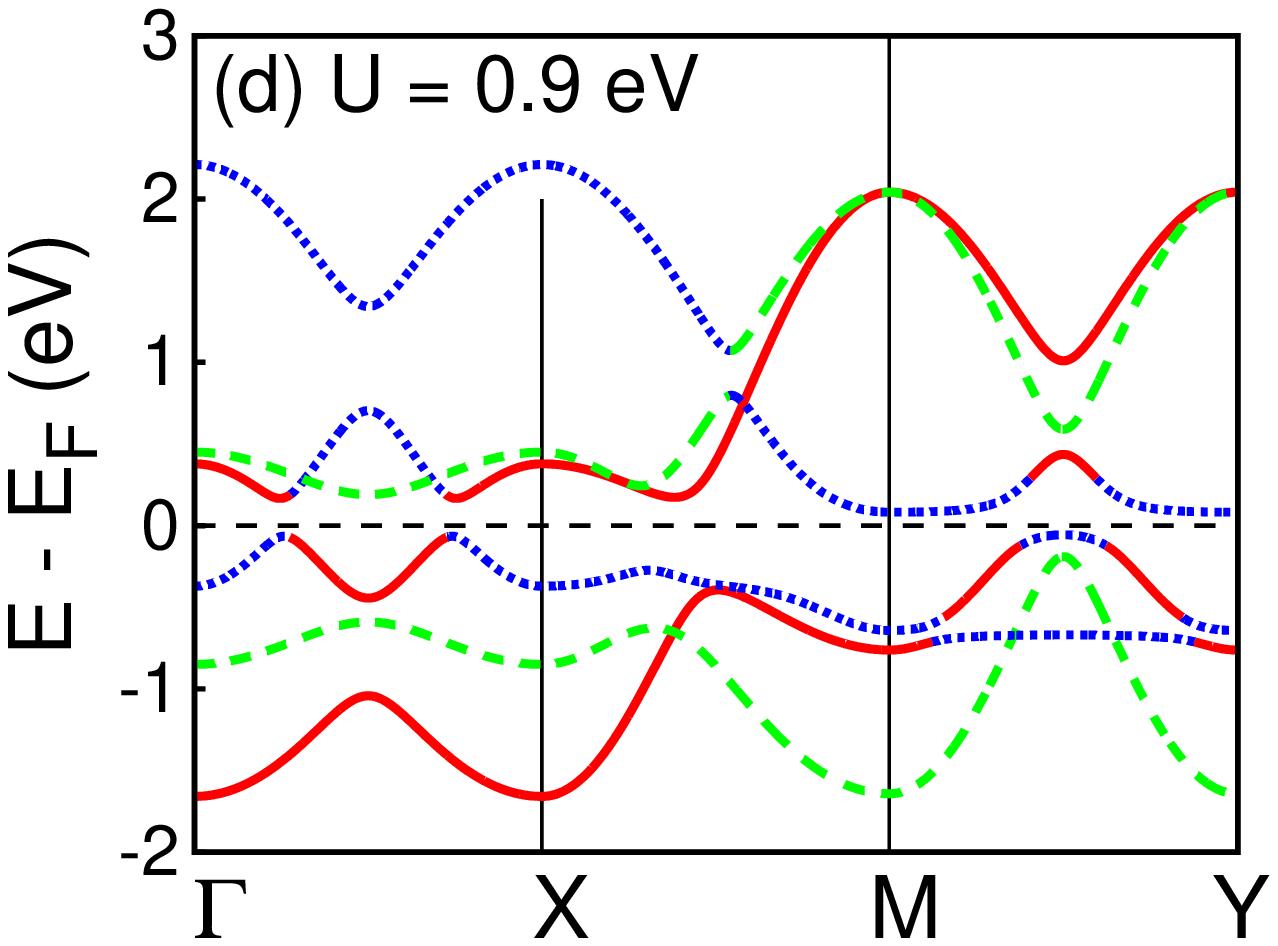}
  \label{dos4}}\\
 \caption{Evolution of the band structure in the $(\pi,0)$ SDW state of the three-band model with increasing interaction strength $U$.}
 \label{dos}
\end{center}
\end{figure}

\begin{figure}
\begin{center}
 \subfigure{\includegraphics[height=70mm]{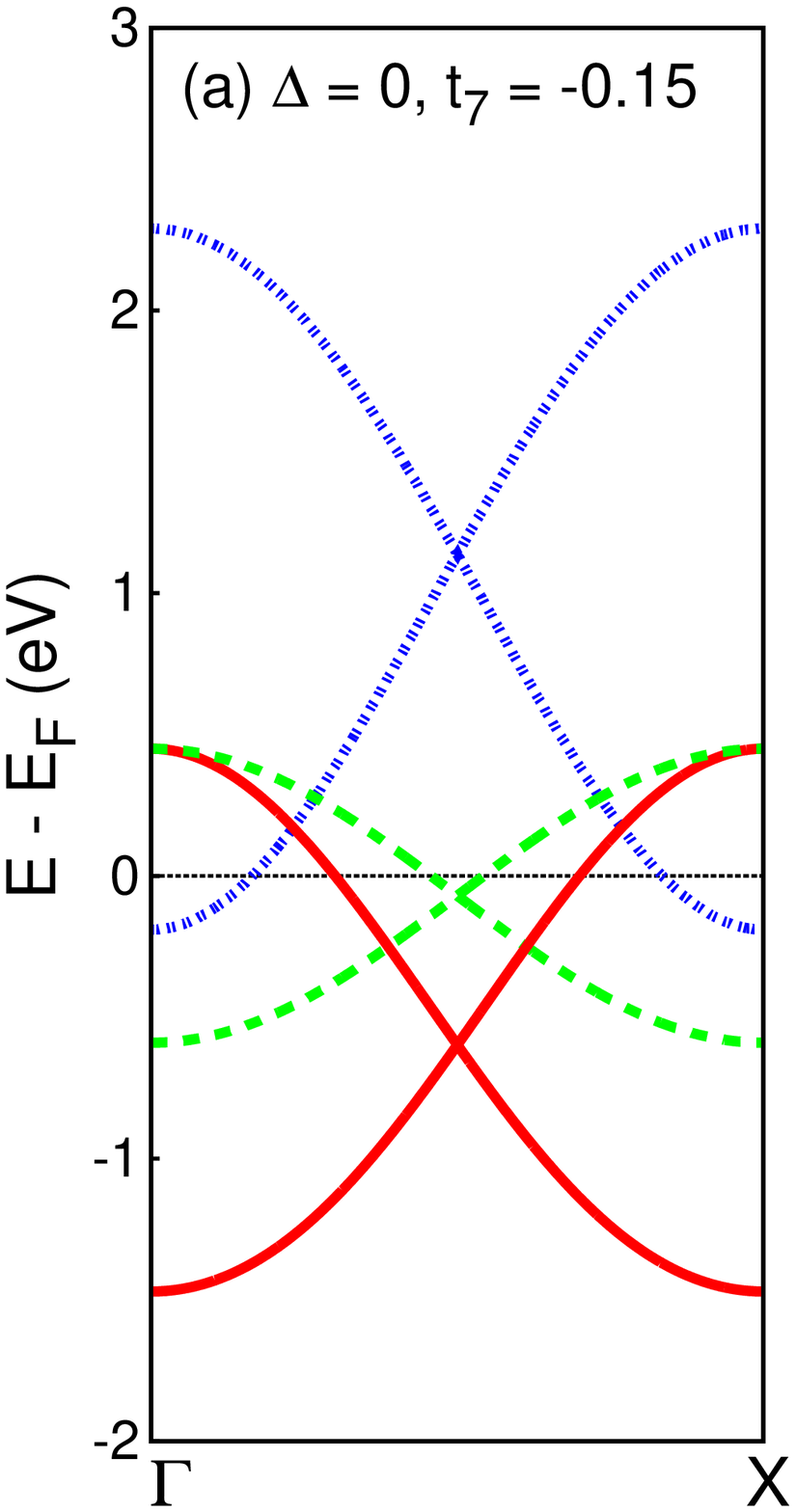}
 \label{bs1}}
 \subfigure{\includegraphics[height=70mm]{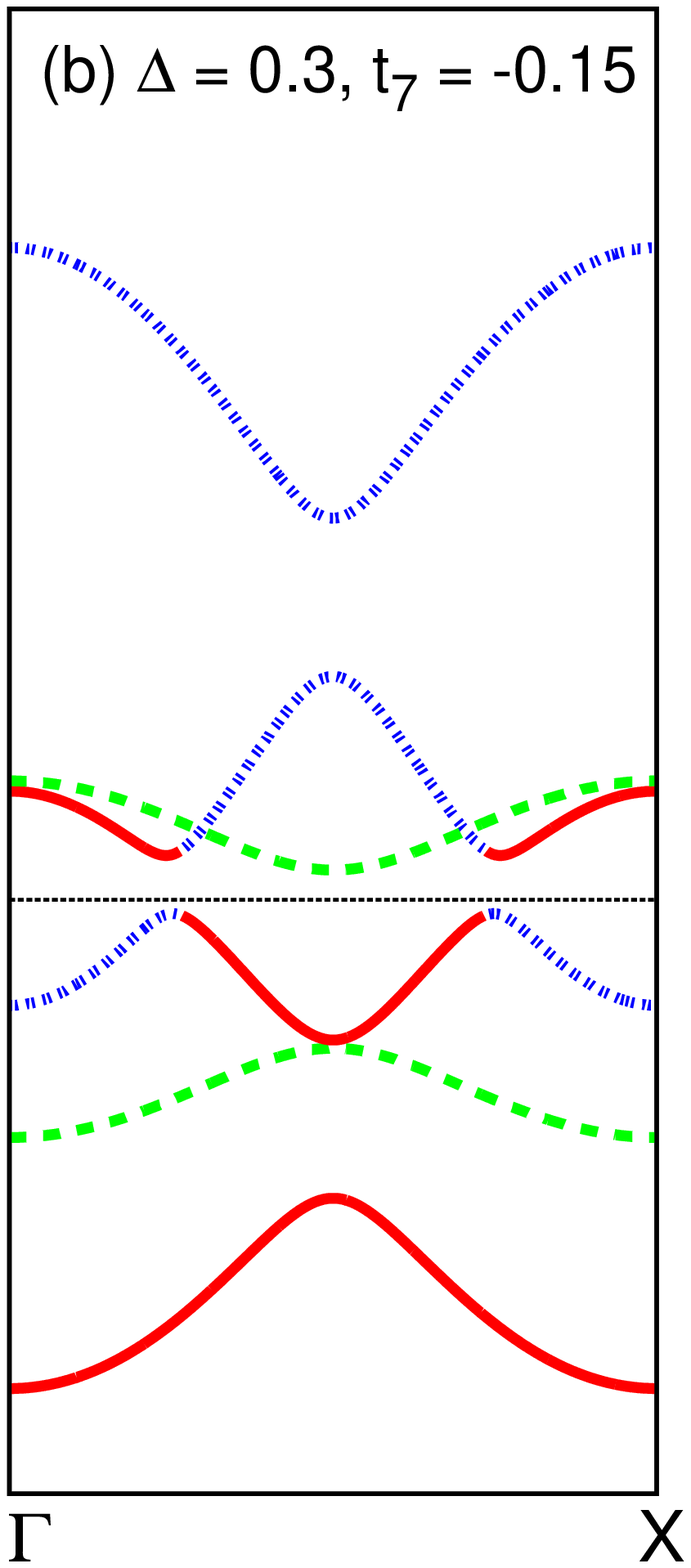}
  \label{bs2}}
   \subfigure{\includegraphics[height=70mm]{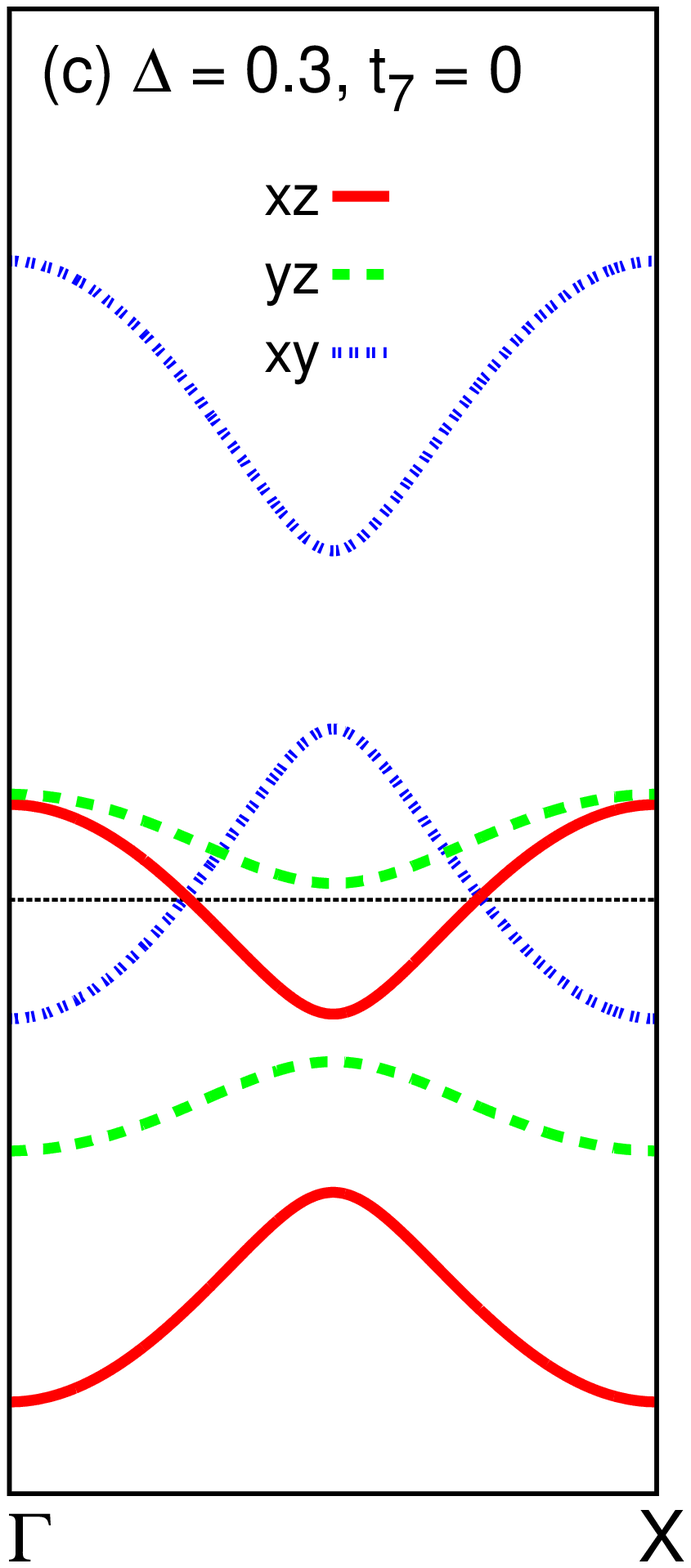}
  \label{bs3}}
 \caption{Band dispersion in the $(\pi,0)$ SDW state along the AF direction ($\Gamma\rightarrow X$), showing that the energy gap opens only when both $\Delta$ and $t_7$ are non-zero. The Fermi energy corresponds to half filling, for which the FS in the PM state of the three-band model is shown in Fig. \ref{fs}.}
 \label{bs}
\end{center}
\end{figure}

The composite effect of the orbital mixing between the $xz$ and $xy$ orbitals and the ($\pi,0$) SDW order on opening the weak band gap near the Fermi energy can be clearly seen by going over to a reduced basis involving the relevant $xz$ and $xy$ bands and contracting over the sublattice basis to obtain the band mixing terms in the magnetic state.

We focus on the $k_x$ direction of the Brillouin zone which is the AF ordering direction, and set $k_y=0$. As the $yz$ orbital completely decouples in this case, we start with the $[4 \times 4]$ part of the Hamiltonian matrix given in Eq. (5) involving the (A$xz$ B$xz$ A$xy$ B$xy$) part of the orbital-sublattice basis for the $xz$ and $xy$ orbitals:

\begin{eqnarray}
\fl
H_{\rm HF}^{\sigma} (k_x,0) =  
\left [ \begin{array}{cccc}
-\sigma \Delta_{xz} - 2 t_2 & \varepsilon_{{\bf k},xz} & 0 & t_{{\bf k},\rm mix}
\\ 
\varepsilon_{{\bf k},xz} & \sigma \Delta_{xz} - 2 t_2 & t_{{\bf k},\rm mix}  &
0 \\ 
0 & t_{{\bf k},\rm mix}^{\ast} & - \sigma \Delta_{xy} -2 t_5 + \varepsilon_{\rm diff} & \varepsilon_{{\bf k},xy} \\ 
t_{{\bf k},\rm mix}^{\ast} & 0 & \varepsilon_{{\bf k},xy} & \sigma \Delta_{xy} -2 t_5 + \varepsilon_{\rm diff}\\
\end{array}
\right ] \nonumber \\ 
\label{Hamiltonian2}
\end{eqnarray}
where $t_{{\bf k},\rm mix} = - 2 i (t_7 + 2 t_8) \sin k_x$ is the mixing term between $xz$ and $xy$ orbitals on opposite sublattices, and the bare band energy terms are $\varepsilon_{{\bf k},xz} = (-2t_1-4t_3) \cos k_x$ and $\varepsilon_{{\bf k},xy} = (-2t_5-4t_6) \cos k_x$.

In the absence of the above mixing term $t_{{\bf k},\rm mix}$, the $xz$ and $xy$ orbital sectors are decoupled and in the ($\pi,0$) ordered SDW state, the SDW band energies for the upper (+) and lower (-) bands are given by:
\begin{eqnarray}
 E_{{\bf k},xz}^{\pm} = - 2t_2 \pm \sqrt{\Delta_{xz}^2 + \varepsilon_{{\bf k},xz}^2} \nonumber \\
 E_{{\bf k},xy}^{\pm} = - 2t_5 +  \varepsilon_{\rm diff} \pm \sqrt{\Delta_{xy}^2 + \varepsilon_{{\bf k},xy}^2}. 
\end{eqnarray}
The exchange fields $\Delta_{xz}$ and $\Delta_{xy}$ open band gaps at $k_x = \pi/2$ where $\varepsilon_{{\bf k},xz} = \varepsilon_{{\bf k},xy} = 0$. However, these gaps open at energies away from the Fermi energy, as discussed earlier [see  Fig. 4]. For $k_x$ close to zero, which is far away from where the band gaps opens, the band energies are nearly unchanged for small $\Delta$ [see Fig. 4(c)], and are given by:
\begin{eqnarray}
 E_{{\bf k},xz}^{\pm} = - 2t_2 \pm  \varepsilon_{{\bf k},xz} \nonumber \\
 E_{{\bf k},xy}^{\pm} = - 2t_5 +  \varepsilon_{\rm diff} \pm  \varepsilon_{{\bf k},xy}. 
\end{eqnarray}

Out of these four bands, we focus on the two relevant SDW bands 
$E_{{\bf k},xz}^{+}$ (upper $xz$ band) and $E_{{\bf k},xy}^{-}$ (lower $xy$ band) which lie near the Fermi energy. We now consider an effective $[2 \times 2]$ Hamiltonian matrix in the reduced basis consisting of these two eigenstates $|{\bf k}_{xz}^+ \rangle$ and $|{\bf k}_{xy}^- \rangle$. The band mixing term in this reduced basis is obtained by evaluating the matrix elements using the corresponding eigenvectors in the two-sublattice basis: 
\begin{eqnarray}
|{\bf k}_{xz}^{+}\rangle = 
\left ( \begin{array}{c} 
\alpha_{{\bf k},xz}^{+} \\ \beta_{{\bf k},xz}^{+}
\end{array} \right ) \;\;\;\;\;\;\;
{\rm and} \;\;\;\;\;\;\;
|{\bf k}_{xy}^{-}\rangle = 
\left ( \begin{array}{c} 
\alpha_{{\bf k},xy}^{+} \\ \beta_{{\bf k},xy}^{+}
\end{array} \right ) 
\end{eqnarray}
for the $xz$ and $xy$ band sectors of the Hamiltonian matrix in Eq. (8). This yields
\begin{eqnarray}
H_{\rm HF}^{\sigma} (k_x,0) =  
\left [ \begin{array}{lr}
E_{{\bf k},xz}^{+} & \delta_{\bf k}^{\rm mo} \\ 
\delta_{\bf k}^{{\rm mo}\ast} & E_{{\bf k},xy}^{-} \\ 
\end{array}
\right ] \nonumber \\ 
\label{Hamiltonian3}
\end{eqnarray}
where 
\begin{equation}
\delta_{\bf k}^{\rm mo} = t_{{\bf k},\rm mix} 
\langle {\bf k}_{xz}^{+} | \sigma_x | {\bf k}_{xy}^{-}\rangle = 
t_{{\bf k},\rm mix} \big( \alpha_{{\bf k},xz}^+ \beta_{{\bf k},xy}^- + \beta_{{\bf k},xz}^+ \alpha_{{\bf k},xy}^-)  
\end{equation}
is a magneto-orbital coupling term involving the orbital mixing term and the SDW state matrix element of $\sigma_x$ representing the opposite sublattice operator. Here $\alpha_{\bf k}$ and $\beta_{\bf k}$ are the $A$ and $B$ sublattice components of the eigenvectors for the two orbitals:
\begin{eqnarray}
\fl
\alpha_{{\bf k},xz}^+ = \sqrt{\frac{1}{2} \Bigg( 1 -\frac{\Delta_{xz}}{\sqrt{\Delta_{xz}^2 + \varepsilon_{{\bf k},xz}^2}}\Bigg)};\;\;\;\; \beta_{{\bf k},xz}^+ = \sqrt{\frac{1}{2} \Bigg( 1+\frac{\Delta_{xz}}{\sqrt{\Delta_{xz}^2 + \varepsilon_{{\bf k},xz}^2}}\Bigg)} \nonumber \\
\fl
\alpha_{{\bf k},xy}^- = \sqrt{\frac{1}{2} \Bigg( 1+\frac{\Delta_{xy}}{\sqrt{\Delta_{xy}^2 + \varepsilon_{{\bf k},xy}^2}}\Bigg)};\;\;\;\; \beta_{{\bf k},xy}^- = - \sqrt{\frac{1}{2} \Bigg( 1-\frac{\Delta_{xy}}{\sqrt{\Delta_{xy}^2 + \varepsilon_{{\bf k},xy}^2}}\Bigg)}
\end{eqnarray}
We note here that the majority/minority amplitudes on the two 
sublattices get exchanged for the $xz$ upper band and $xy$ lower band.

In the absence of any SDW order ($\Delta_{xz}=\Delta_{xy}=0$), the magneto-orbital coupling term $\delta_{\bf k}^{\rm mo}=0$ as the A and B sublattice amplitudes are identical in magnitude ($|\alpha_{\bf k}|=|\beta_{\bf k}|=1/\sqrt{2}$), so the negative sign in Eq. (12) yields an exact cancellation. However, for finite $\Delta_{xz}$ and $\Delta_{xy}$, the sublattice asymmetry in the magnetic state results in $\delta_{\bf k}^{\rm mo} \neq 0$, which leads to strong mixing at $\bf k$-points where the two band energies $E_{{\bf k},xz}^{+}$ and $E_{{\bf k},xy}^{-}$ are nearly degenerate, resulting in the opening of the non-conventional weak energy gap near the Fermi energy. For $\varepsilon_{{\bf k},xz}^2 \simeq \varepsilon_{{\bf k},xy}^2 \simeq \varepsilon_{\bf k}^2$ and $\Delta_{xz} \simeq \Delta_{xy} \simeq \Delta$, we obtain:
\begin{equation}
\delta_{\bf k}^{\rm mo}=[2i(t_7+2t_8) \sin k_x ] 
\left [ \frac{\Delta}{\sqrt{\Delta^2+\varepsilon_{\bf k}^2}} \right ],
\end{equation}
explicitly showing the composite dependence of the magneto-orbital coupling term on {\it both} the orbital mixing and SDW order, and thus accounting for the absence of the band gap near the Fermi energy if either of these two terms vanishes, as seen in Fig. 4.

\section{Conclusions}
A magneto-orbital coupling mechanism was proposed and shown to account for the opening of a non-conventional weak energy gap near the Fermi energy in the $(\pi,0)$ SDW state of a realistic three band model for iron pnictides at half filling. The coupling term involves a composite dependence on both the orbital mixing (hybridization) terms between the $xy$ and $xz/yz$ orbitals as well as the SDW order parameter, thus accounting for the absence of the energy gap if either of these two terms is absent. As the orbital mixing terms also play an important role in the mixed composition ($xy$ and $xz/yz$) of the elliptical electron pockets in the non-magnetic-state FS structure, the magneto-orbital coupling provides a subtle link between the non-magnetic-state FS features and the SDW state weak energy gap. Furthermore, the non-conventional nature of the weak energy gap for quasiparticle  excitations with orbital mixing provides a sharp contrast from the conventional strong SDW gap in cuprate antiferromagnets.   

\section*{Acknowledgements}
SG and NR acknowledge financial support from Council of Scientific and Industrial Research, India.

\end{document}